\documentclass[12pt]{article}
\usepackage[dvips]{graphics}
\def\NPB#1#2#3{{Nucl.~Phys.} {\bf{B#1}} (19#2) #3}
\def\PLB#1#2#3{{Phys.~Lett.} {\bf{B#1}} (19#2) #3}
\def\PRD#1#2#3{{Phys.~Rev.} {\bf{D#1}} (19#2) #3}
\def\PRL#1#2#3{{Phys.~Rev.~Lett.} {\bf#1} (19#2) #3}
\def\DROP#1{\makebox[-1.2mm][r]{$\displaystyle{#1}$}}
\def\mean#1{\left<{#1}\right>}
\def\pfrac#1#2{\left(\frac{#1}{#2}\right)}
\def\GeV{{\rm GeV}}

\def\sec{{\rm sec}}

\makeatletter
\def\gsim{\Compoundrel>\over\sim}
\def\lsim{\Compoundrel<\over\sim}
\def\Compoundrel#1\over#2{\mathpalette\CompoundreL{{#1}\over{#2}}}
\def\CompoundreL#1#2{\CompoundREL#1#2}
\def\CompoundREL#1#2\over#3{\mathrel
  {\vcenter{\hbox{$\m@th\buildrel{#1#2}\over{#1#3}$}}}}
\makeatother
\begin{document}
\begin{titlepage}
\vspace{3cm}
\begin{flushright}
TU-539, RCNS-98-05 \\
May 1998
\end{flushright}
\vskip 1cm
\begin{center}
{\Large \bf Late-Time Entropy Production and \\
        Relic Abundances of Neutralinos}
\end{center}
\vskip 1.5cm  
\begin{center}
{\large Takaaki Nagano and Masahiro Yamaguchi} \\
\vskip 1.5cm
{\em Department of Physics, Tohoku University, \\
Sendai 980-8578, Japan}
\end{center}
\vskip 2cm
\thispagestyle{empty}
\begin{abstract}
Many models possess unwanted relics, which should be diluted by entropy 
production just before the big-bang nucleosynthesis.
A field responsible for the
entropy production  may produce stable weakly interacting
massive particles, if kinematically accessible. We compute their
relic abundances by integrating out coupled equations numerically. Applying
our results to supersymmetric standard models, we argue that the 
neutralino lightest superparticles will overclose the universe in most of
the parameter space if the reheat temperature is of the order of 10 MeV.

\end{abstract}
\end{titlepage}
Many models beyond the standard model predict unwanted relics, which
should be diluted by entropy production such as inflation and
subsequent reheating.  Among other things, weakly interacting massive
scalar fields will be the most problematic, as they can store their
energy density in the form of coherent oscillation, which may not be
diluted by inflation because the coherent mode during the inflationary
epoch is generally displaced from its value in the true vacuum.  An
example is a problem associated with moduli fields of string theory,
named the cosmological moduli problem \cite{moduli}.  In string theory
coupling constants are determined dynamically by vacuum expectation
values of the moduli (and dilaton) fields. These fields generically
acquire masses comparable to the gravitino mass, which should be in
the TeV range in gravity mediated supersymmetry breaking.  Damped
coherent oscillation and subsequent decay of the fields would spoil
the standard big-bang nucleosynthesis (BBN). Late-time entropy
production before the BBN should operate to dilute the energy density
of the problematic scalar fields. Indeed thermal inflation was
proposed \cite{LS,SKY}, for which reheat temperature should be much
lower than the electroweak scale for a successful dilution.

Another example of the problematic scalar fields is an invisible axion
with decay constant larger than $10^{13}$ GeV, whose coherent
oscillation would exceed the critical density of the universe 
\cite{axion-bound}. Entropy production may be able to dilute the energy 
density of the axion field \cite{axion-entropy}. 
Since
the oscillation of the axion field starts around the QCD phase
transition, the reheat temperature at the entropy production should be
lower than 100 MeV. Interestingly it has been argued that strongly coupled
$E_8 \times E_8$ heterotic string theory (M-theory) possesses such an
axion candidate with the decay constant $\sim 10^{16}$ GeV, provided
that non-perturbative effects to generate the superpotential for the
axion field are suppressed \cite{BD,Choi}. To dilute
the energy density of the M-theory axion to a harmless level, the
reheat temperature of the entropy production should be at its lowest
value allowed by the BBN, 1--10 MeV.

The requisite entropy production is provided by decays of massive
unstable particles.  The decays, however, may produce too many stable
particles as well if the production is kinematically allowed. In a
supersymmetric theory, the lightest superparticle (LSP), which is likely to
be a neutralino, is stable under the assumption of $R$-parity
conservation.  Indeed it was argued that the neutralinos produced by
the decays of the massive particles tend to overclose the universe, if
the reheat temperature is much lower than  1 GeV \cite{MYY,KMY}.

Motivated by the aforementioned demands for the late-time entropy
production mechanism, we shall reexamine the problem of the relic
abundances of the neutralinos which are produced by the decays of
massive unstable particles.  Given an annihilation cross section of
the neutralinos, we will solve coupled equations numerically to
evaluate the relic abundances, taking into account effects of
exponential decays of the unstable particles, whereas an approximation
of simultaneous decay was made in the previous works \cite{MYY,KMY}.
It turns out that results obtained in our numerical calculation agree
with the previous estimates up to a numerical factor of order unity
and hence we confirm the validity of arguments in
Refs.~\cite{MYY,KMY}.  Note that our computation should be applicable
to any stable weakly-interacting-massive-particles. We will then argue
that the stable LSP suffers from the overclosure problem in most of
the parameter space if the reheat temperature is of the order of 10
MeV or less.  We will also briefly mention possible implications of
our analysis when there remains the M-theory axion in the massless
spectrum of the heterotic M-theory.

We consider the situation that massive unstable particles $\phi$ of the 
decay width $\Gamma$ come to dominate the energy density of the universe, 
then decay to neutralinos $\chi$ as well as to radiation with branching 
ratios $B$ and $1-B$, respectively.  
By assuming two-body decay, the energy density of $\chi$ just after
the  production is $m_{\phi}/2$.  They subsequently lose their energy, 
by the scattering with thermal bath, down to $\simeq m_{\chi}\,(\gg T)$.  
Assuming this energy transportation occur rapidly, we define the effective 
branching ratio $B_{\rm eff} \equiv B \frac{m_{\chi}}{m_{\phi}/2}$.  

Time evolution of cosmic scale factor $R$ and 
energy densities $\rho_{\phi}$, $\rho_R$, $\rho_{\chi}$ of $\phi$, 
radiation, and $\chi$, respectively, are
described by the following Friedmann equation and (integrated) 
Boltzmann equations: 
\begin{eqnarray}
  &&
  H^2 = \pfrac{\dot{R}}{R}^2
  = \frac{1}{3M^2} (\rho_{\phi}+\rho_{R}+\rho_{\chi}) ~,
  \\
  &&
  \frac{d}{dt} (R^3 \rho_{\phi}) 
  = - \Gamma (R^3 \rho_{\phi}) ~,
  \\
  &&
  \frac{d}{dt} (R^3 \rho_{R}) + \frac{\rho_R}{3} \frac{d R^3}{dt}
  = (1-B_{\rm eff}) \Gamma (R^3 \rho_{\phi})
  + \frac{\mean{\sigma v}}{m_{\chi}} (R^3 \rho_{\chi})^2 \frac{1}{R^3} ~,
  \\
  &&
  \frac{d}{dt} (R^3 \rho_{\chi}) 
  = B_{\rm eff} \Gamma (R^3 \rho_{\phi})
  - \frac{\mean{\sigma v}}{m_{\chi}} (R^3 \rho_{\chi})^2 \frac{1}{R^3} ~,
\end{eqnarray}
where $M$ is the reduced Planck mass: it is related to the Newton's
constant as $G = 1/m_{\rm Pl}^2 = 1/(8\pi M^2)$, and $\mean{\sigma v}$
is an ensemble-averaged annihilation cross section of neutralinos
times their relative velocity.  We treat $\mean{\sigma v}$ as a
constant, which is indeed the case when the $S$-wave is dominant. When a
higher partial wave dominates, on the other hand, $\mean{\sigma v}$
should be interpreted as a sort of weighted average in the time
interval around when the annihilation process is important. In the
above equations, we did not include pair production of the
neutralinos, which is not effective at low temperatures.  Note that
the right-hand side of the Boltzmann equation for $\rho_R$ is
consistent with the energy conservation within comoving volume for the
case $\dot{R} = 0$.  Following~\cite{ST} we introduce the
dimensionless variables:
\begin{eqnarray}
  &&
  x \equiv \Gamma t ~,\quad
  z \equiv R/R_i ~,\quad
  \\
  &&
  f_{\phi} \equiv \rho_{\phi}/\rho_{\phi i} ~,\quad
  f_{R}    \equiv \rho_{R}/\rho_{\phi i} ~,\quad
  f_{\chi} \equiv \rho_{\chi}/\rho_{\phi i} ~,\quad
  \\
  &&
  F_{\phi} \equiv z^3 f_{\phi} ~,\quad
  F_{R}    \equiv z^4 f_{R}    ~,\quad
  F_{\chi} \equiv z^3 f_{\chi} ~,\quad
  \\
  &&
  x_H \equiv \Gamma H_i^{-1}
  = \Gamma
    \left[ \frac{\rho_{\phi i}}{3M^2} (1+f_{Ri}+f_{\chi i}) \right]^{-1/2}
  \\
  &&
  \rho_{\chi 0} \equiv \frac{m_{\chi}\Gamma}{\mean{\sigma v}} ~,\quad
  f_{\chi 0} \equiv \rho_{\chi 0}/\rho_{\phi i} ~.
\end{eqnarray}
Here subscripts $i$ mean their initial values.  
By using these variables, 
the coupled differential equations above can be written as follows: 
\begin{eqnarray}
  &&
  \DROP{\frac{z'}{z}}
  =
  x_H^{-1} (1+f_{Ri}+f_{\chi i})^{-1/2}
  (z^{-3} F_{\phi} + z^{-4} F_R + z^{-3} F_{\chi})^{1/2} ~,
  \\
  &&
  \DROP{F_{\phi}'}
  =
  - F_{\phi} ~,
  \\
  &&
  \DROP{F_R'}
  =
  (1-B_{\rm eff}) z F_{\phi} 
  + f_{\chi 0}^{-1} z^{-2} F_{\chi}^2 ~,
  \\
  &&
  \DROP{F_{\chi}'}
  =
  B_{\rm eff} F_{\phi}
  - f_{\chi 0}^{-1} z^{-3} F_{\chi}^2 ~, 
\end{eqnarray}
where prime denotes $d/dx$.  
We impose the following initial conditions at $x=x_i$, 
\begin{eqnarray}
  &&
  \qquad\quad
  z_i = 1 ~,\quad
  F_{\phi i} = 1 ~,
  \\
  &&
  F_{R i} = \frac{\rho_{Ri}}{\rho_{\phi i}} \,(\ll 1) ~,\quad
  F_{\chi i} = \frac{\rho_{\chi i}}{\rho_{\phi i}} \,(\ll 1) ~.
\end{eqnarray}

There is no difficulty in solving the Boltzmann equation for $F_{\phi}$:
\begin{eqnarray}
  F_{\phi} = e^{-(x-x_i)} ~.
\end{eqnarray}

By assuming $\rho \propto R^{-3}$ for $x_i \lsim x \lsim 1$, 
\begin{eqnarray}
  z^{3/2} = \frac{3}{2} \frac{x}{x_H}
  \quad\hbox{and}\quad
  \Gamma H^{-1} = \frac{3}{2} x
\end{eqnarray}
for $x_i \ll x \lsim 1$.  
Our numerical calculation shows that $z|_{x=1}$ is about $10\%$ larger than 
the estimate $(\frac{3}{2}x_H^{-1})^{2/3}$.  

With an approximation of the simultaneous $\phi$ decay at $x=1$, one obtains 
\begin{eqnarray}
  &&
  \DROP{F_R|_{x\gg 1}}
  =
  z|_{x=1} \cdot F_{\phi}|_{x\ll 1}
  \nonumber \\
  &&
  \simeq
  \left( \frac{3}{2}x_H^{-1} \right)^{2/3}
  \nonumber \\
  &&
  =
  1.20 \times (1.09\,x_H^{-2/3}) ~.
\end{eqnarray}
Numerical calculation~\cite{ST} for the coupled equations
shows, on the other hand,
\begin{eqnarray}
  &&
  \DROP{F_R|_{x\gg 1}}
  =
  \int_0^\infty\!z e^{-x}\,dx
  \nonumber \\
  &&
  =
  1.09\,x_H^{-2/3} ~,
 \label{eq:F-R:numerical}
\end{eqnarray}
for $F_{Ri} < 10^{-3}$ and $B_{\rm eff} < 10^{-3}$.  
We confirm that the formula above is correct also for larger $F_{Ri}$ 
and/or $B_{\rm eff}$ with at most about $10\%$ error.  

Our main concern in this paper is $F_{\chi}$ for $x\gg1$, to which 
$\rho_{\chi}$ is proportional.  
A crude behavior of $F_{\chi}$ can  be seen as follows.  
If $f_{\chi 0}^{-1} = \frac{\mean{\sigma v}}{m_{\chi}\Gamma} \rho_{\phi i}$ 
is negligibly small, or quantitatively 
if $B_{\rm eff} x_H^2 f_{\chi 0}^{-1} < 1$ as we shall see shortly, 
\begin{eqnarray}
  F_{\chi}
  =
  \left\{
      \begin{array}{lll}
        B_{\rm eff} (x-x_i) + F_{\chi i} & {\rm ,~for} & x_i < x \lsim 1 ~\\
        B_{\rm eff}                      & {\rm ,~for} & x \gsim 1 ~.
      \end{array}
    \right.
\end{eqnarray}
That is, $F_{\chi}$ increases monotonically with $x$.  
On the other hand, if $f_{\chi 0}^{-1}$ is not so small 
the second term of the Boltzmann equation for $F_{\chi}$ cannot be neglected 
when $F_{\chi}$ becomes large: 
\begin{eqnarray}
  &&
  \DROP{\left. F_{\chi}' \right|_{x=1}}
  \simeq
  B_{\rm eff} e^{-1}
  [ 1 - 1.2 B_{\rm eff} x_H^2 f_{\chi 0}^{-1} ] ~.
\end{eqnarray}
Thus when $B_{\rm eff} x_H^2 f_{\chi 0}^{-1} > 1$, 
$F_{\chi}$ increases until $x \sim 1$, but decreases thereafter 
to reach $F_{\chi}|_{x\gg 1} < B_{\rm eff}$.
Note that 
\begin{eqnarray}
  &&
  \DROP{B_{\rm eff} x_H^2 f_{\chi 0}^{-1}}
  =
  6 M^2 B \Gamma \frac{\mean{\sigma v}}{m_{\phi}}
  \nonumber \\
  &&
  =
  1.17
  \pfrac{\mean{\sigma v}}{10^{-10}\GeV^{-2}}
  \pfrac{m_{\phi}}{10^3\GeV}^{-1}
  \pfrac{B}{0.5}
  \pfrac{\Gamma^{-1}}{1\sec}^{-1}
  ~.
\end{eqnarray}

Fig.~1 shows the evolution of $F_{\chi}$, obtained by integrating out
the coupled differential equations numerically. It shows it increases
till $x=2$ or so, then decreases to reach a constant value. 
In the same figure, we also show the evolutions of the other quantities, 
$z$, $F_{\phi}$ and $F_R$.  

Let us now compare our numerical results with a naive estimate 
given in~\cite{MYY}.  
This naive estimation of $F_{\chi}|_{x\gg 1}$ for
the case where the annihilation process becomes effective was obtained
by equating the annihilation rate $\mean{\sigma v}
\rho_{\chi}/m_{\chi}$ to the expansion rate $H$ at the decay epoch ($x=1$) 
of $\phi$, with the approximation of the simultaneous decay. One finds
\begin{eqnarray}
  \rho_{\chi}|_{x=1}
  =
  \left. \frac{m_{\chi}H}{\mean{\sigma v}} \right|_{x=1}
  =
  \frac{2}{3} \frac{m_{\chi}\Gamma}{\mean{\sigma v}} ~,
\end{eqnarray}
or, 
\begin{equation}
  \DROP{F_{\chi}|_{x\gg 1}}
  =
  F_{\chi}|_{x=1}
  =
  \frac{3}{2} x_H^{-2} f_{\chi 0} ~. 
  \label{eq:F-chi:naive}
\end{equation}
On the other hand, we find the following fit for our numerical results
(see Fig.~2): 
\begin{eqnarray}
  &&
  \DROP{\left. F_{\chi} \right|_{x\gg 1}}
  =
  1.52 (B_{\rm eff} x_H^2 f_{\chi 0}^{-1})^{-0.919} B_{\rm eff} 
 \label{eq:F-chi:numerical}
\end{eqnarray}
for $10 < B_{\rm eff} x_H^2 f_{\chi 0}^{-1} < 10^8$ within $15\%$
error.  Comparison of the two results shows that the naive estimate is
correct up to a factor of order unity in a wide range of the parameters.   
As $B_{\rm eff} x_H^2 f_{\chi 0}^{-1}$ $\propto$ 
$B \Gamma \frac{\mean{\sigma v}}{m_{\phi}}$  becomes larger, 
$F_{\chi}|_{x\gg 1}$ is more underestimated if one uses the naive estimation.
For example, 
for $B_{\rm eff} = 10^{-1}$ and $x_H^2 f_{\chi 0}^{-1} = 10^8$, 
\begin{eqnarray}
  \left. F_{\chi} \right|_{x\gg 1}^{\rm naive}
  =
  1.5 \times (10^8)^{-1}
  =
  1.5 \times 10^{-8} ~,
\end{eqnarray}
while, 
\begin{eqnarray}
  &&
  \DROP{\left. F_{\chi} \right|_{x\gg 1}^{\rm numerical}}
  =
  1.52 \times (10^{-1} \times 10^8)^{-0.919} \times 10^{-1}
  \nonumber \\
  &&
  =
  5.61 \times 10^{-8}
  \nonumber \\
  &&
  =
  3.7 \times \left. F_{\chi} \right|_{x\gg 1}^{\rm naive} ~.
\end{eqnarray}

Now we turn to $\rho_{\chi}/s|_{x\gg 1}$ where  $s$ stands for entropy density.
Using 
\begin{eqnarray}
  &&
  \DROP{\rho_R} 
  = a T^4
  ~, \nonumber\\
  &&
  \DROP{s} 
  = \frac{4}{3} \frac{\rho_R}{T}
  = \frac{4}{3} a^{1/4} \rho_R^{3/4}
  ~, \\
  &&
  \DROP{a}
  = \frac{\pi^2}{30} g_*
  \nonumber
\end{eqnarray}
with $g_*$ being the effective relativistic degrees of freedom, we find
\begin{equation}
  \frac{\rho_{\chi}}{s}=\frac{3}{4}
  \frac{\rho_{\phi i} F_{\chi}}{a^{1/4} F_R^{3/4}}
    = \frac{3^{5/4}}{4}
      \frac{F_{\chi}}{a^{1/4}(x_H^{2/3} F_R)^{3/4}} 
      \left( M \Gamma \right)^{1/2} ~.
\end{equation}
Substituting our numerical results 
Eqs.~(\ref{eq:F-R:numerical}) and (\ref{eq:F-chi:numerical}), 
we obtain the following fit
\begin{eqnarray}
  &&
  \DROP{\left.\frac{\rho_{\chi}}{s}\right|_{x\gg 1}}
  =
  \frac{3^{5/4}}{4}
  (1.09)^{-3/4}
  \left( \frac{\pi^2}{30}g_* \right)^{-1/4}
  (M\Gamma)^{1/2}
  \nonumber \\
  &&
  \qquad
  \times
  1.52
  \left( 6 M^2 B \Gamma \frac{\mean{\sigma v}}{m_{\phi}} \right)^{-0.919}
  \times
  2 B \frac{m_{\chi}}{m_{\phi}}
  \nonumber \\
  &&
  =
  3.08 h^{-2} \times 10^4 \left. \frac{\rho_{\rm cr}}{s} \right|_{0}
  \nonumber\\
  &&
  \qquad
  \times
  \pfrac{\mean{\sigma v}}{10^{-10}\GeV^{-2}}^{-0.919}
  \pfrac{m_{\chi}}{10^2\GeV}
  \pfrac{\Gamma^{-1}}{1\sec}^{0.419}
  \nonumber\\
  &&
  \qquad
  \times
  \pfrac{m_{\phi}}{10^3\GeV}^{-0.081}
  \pfrac{B}{0.5}^{0.081}
  \pfrac{g_*}{10.75}^{-0.46} ~, 
\end{eqnarray}
where $\rho_{\rm cr}/s|_0 = 3.64h^2\times 10^{-9}\GeV$ is the ratio of
the critical density $\rho_{\rm cr}$ to the entropy density $s$ at
present.  This formula reproduces values of $\rho_{\chi}/s|_{x\gg 1}$
calculated numerically within $15\%$ error for $10 < B_{\rm eff} x_H^2
f_{\chi 0}^{-1} < 10^8$. Alternatively, the annihilation cross section
is related to the relic abundance as
\begin{eqnarray}
  &&
  \DROP{\mean{\sigma v}}
  =
  0.767h^{-2.18}\times10^{-5}\GeV^{-2}
  \nonumber\\
  &&
  \qquad
  \times
  \pfrac{\rho_{\chi}/s|_{x\gg 1}}{\rho_{\rm cr}/s|_0}^{-1.09}
  \pfrac{m_{\chi}}{10^2\GeV}^{1.09}
  \pfrac{\Gamma^{-1}}{1\sec}^{0.456}
  \nonumber\\
  &&
  \qquad
  \times
  \pfrac{m_{\phi}}{10^3\GeV}^{-0.0881}
  \pfrac{B}{0.5}^{0.0881}
  \pfrac{g_{*}}{10.75}^{-0.501}.
  \label{eq:sigma-v}
\end{eqnarray}
One finds that Eq.~(\ref{eq:sigma-v}) agrees with the previous
estimate in Ref.~\cite{MYY}, up to a factor of order unity, if one
recalls the temperature when $\phi$ decays is 
\begin{equation}
  T 
  \simeq
  g_*^{-1/4} \sqrt{M\Gamma}
  \sim 
  \left(\frac{\Gamma^{-1}}{1 \mbox{sec}} \right)^{-1/2} \mbox{MeV} ~,
\end{equation}
which may be referred to as the reheat temperature.

The ensemble-averaged annihilation cross section (times relative velocity) 
$\mean{\sigma v}$ can be, in general, expanded in the averaged velocity 
squared  $\mean{v^2}$ of the LSP:
\begin{equation}
 \mean{\sigma v}=a + b \mean{v^2} + \cdots ~.
\end{equation}
The coefficients $a$, $b$, {\it etc.}\ strongly depend on the
composition of the LSP as well as superparticle mass spectra \cite{DN}.
Sometimes the coefficient $a$ of the $S$-wave becomes very small due
to the Majorana nature of the neutralino. This is indeed the case when
the LSP is the bino-dominant neutralino, in which $a$ is proportional
to the mass squared of the fermion in the final state.  Then, in order
to evaluate $\mean{\sigma v}$ and subsequently $\Omega_{\chi} h^2$, we
have to know $\mean{v^2}$ precisely. The authors of Ref.~\cite{KMY}
found that the bino-dominant LSP is likely to reach kinetic
equilibrium and thus $\mean{v^2} \simeq 3T/m_{\chi}$ when the reheat
temperature $T$ is somewhat high, $T \gsim$ a few $\times$ 10 MeV.  Using
this value of $\mean{v^2}$, they concluded that the LSP would
overclose the universe as far as the reheat temperature is lower than
a few hundred MeV. It is not clear, however, whether the LSP is in the
equilibrium or not for other sets of parameters.  Without detailed
knowledge of the velocity distribution, there are still some
conclusions we can draw in general.  For instance, in view of the fact
that the annihilation cross section is generally bounded from above as
\begin{equation}
   \mean{\sigma v} \lsim \frac{\pi \alpha_2^2}{m_{\chi}^2} 
                  \sim \frac{10^{-2}}{m_{\chi}^2}
\label{eq:cross-bound}
\end{equation}
we may conclude that the LSP with the mass at a few tens GeV or more
does not annihilate sufficiently enough to survive the overclosure limit,
if the reheat temperature is relatively low, say $T \lsim {\cal O}(10)$
MeV, or $\Gamma^{-1} \gsim {\cal O}(10^{-4})$ sec, which seems to be
required to make the M-theory axion viable.  Note that the bound
(\ref{eq:cross-bound}) does not apply when $s$-channel resonances 
(by $Z$ boson or Higgs bosons) occur. To discuss the relic abundance in this
case, it seems that we need to evaluate the velocity distribution of the LSP,
which is beyond the scope of the present paper.

Finally we would like to discuss implications of our analysis to the
M-theory if it turns out that the axion with large decay constant
indeed arises.  One needs to invoke an entropy production mechanism to
dilute the energy density of the axion's coherent oscillation, unless
its initial amplitude can be tuned to be very small.  As we argued,
the entropy production would produce too many neutralinos, if
kinematically allowed, which would not annihilate efficiently. As a
result they would overclose the universe, if they remain today. This
suggests that R-parity should be violated, making the neutralino-LSP
unstable. Of course there are many loopholes to this argument, and we
would not insist that the R-parity must be broken in the heterotic
M-theory. Rather we regard the R-parity violation as in interesting
option to which more attention should be paid.

\section*{Acknowledgments}
This work was  supported in part by 
the Grant--in--Aid for Scientific Research from the Ministry of 
Education, Science and Culture of Japan No.\ 09640333.

\subsection*{Figure captions}
\begin{description}
\item[Figure 1] Time evolution ($x \equiv \Gamma t$) of 
  $z \equiv R/R_i$, $F_{\phi} \equiv z^3 f_{\phi}$, 
  $F_R \equiv z^4 f_R$, $F_{\chi} \equiv z^3 f_{\chi}$, 
  and $\rho_{\chi}/s$, 
  where 
  $f_{*} \equiv \rho_{*}/\rho_{\phi i}$ ($*$ $=$ $\phi$, $R$, $\chi$) 
  and 
  $\Gamma$ is the decay width of $\phi$, 
  for the case when 
  $B_{\rm eff} \equiv 2 m_{\chi}/m_{\phi} = 10^{-1}$, 
  $x_H \equiv \Gamma H_i^{-1} = 10^{-2}$, 
  $f_{\chi 0} \equiv m_{\chi}\Gamma/(\mean{\sigma v} \rho_{\phi 0}) 
  = 10^{-9}$, 
  $\Gamma^{-1} = 1\sec$, 
  and 
  $g_* = 10.75$.  
  We can see that entropy $S \propto F_R^{3/4}$ within comoving volume 
  increases during $0.01 \lsim x \lsim 2$.  
  We can also see $F_{\chi}$ increases until $x\simeq 2$ but decreases 
  thereafter to reach certain constant value.  
  Likewise, $\rho_{\chi}/s$ decreases for $x \gsim 2$ to reach certain 
  asymptotic value.  
\item[Figure 2] Asymptotic values of $F_{\chi}$ 
  and $\rho_{\chi}/s$ as a function of $f_{\chi 0}^{-1}$, 
  for the case $B_{\rm eff} = 10^{-1}$, $x_H = 10^{-2}$, 
  $\Gamma^{-1} = 1\sec$, and $g_* = 10.75$.  
  Qualitative behavior of $F_{\chi}|_{x \gg 1}$ and 
  $\rho_{\chi}/s|_{x \gg 1}$ changes depending on if 
  $B_{\rm eff} x_H^2 f_{\chi 0}^{-1}$ is greater than $1$.  
  We also show naive estimations, 
  $F_{\chi}|_{x \gg 1}^{\rm naive} = (3/2) x_H^{-2} f_{\chi 0}$ and 
  $\rho_{\chi}/s|_{x \gg 1}^{\rm naive} = 
   2^{-3/2} 3^{3/4} a^{-1/4} \sqrt{M\Gamma}\ F_{\chi}|_{x \gg 1}^{\rm naive}$, 
  and linear fits for $10 < B_{\rm eff} x_H^2 f_{\chi 0}^{-1} < 10^8$, 
  for comparison.  
  Linear fits are accurate within $15\%$ error.  
\end{description}

\begin{figure}[htbp]
  \vspace*{-7cm}
  \hspace*{-2cm}
  \includegraphics{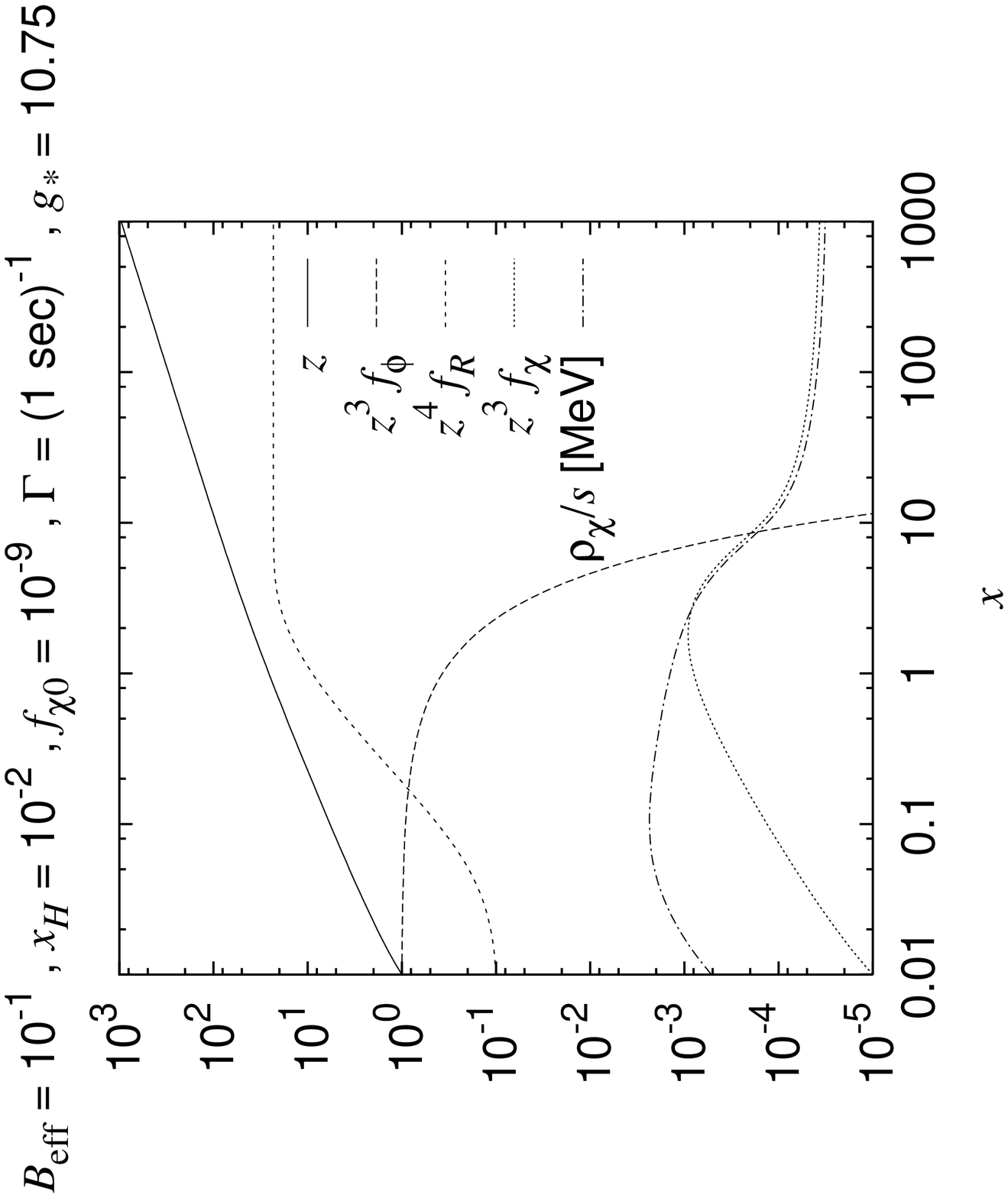}
  \vspace*{-2cm}
  \begin{center}
    Figure~1
  \end{center}
\end{figure}

\begin{figure}[htbp]
  \vspace*{-7cm}
  \hspace*{-2cm}
  \includegraphics{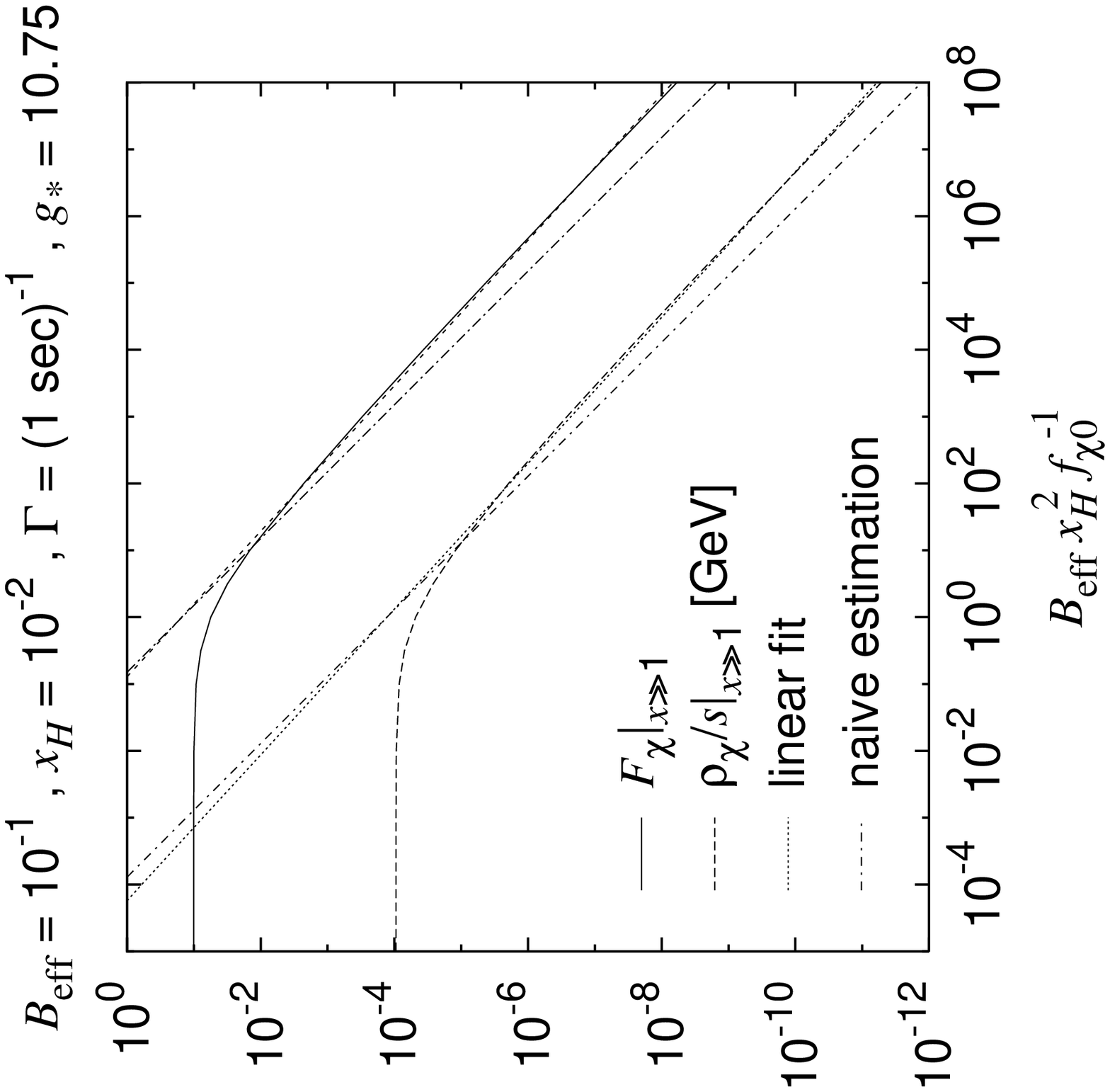}
  \vspace*{-2cm}
  \begin{center}
    Figure~2
  \end{center}
\end{figure}
\end{document}